\documentclass[10pt]{article}

\usepackage{amsmath}
\usepackage{amssymb}

\usepackage{graphicx}
\usepackage{cite}
\usepackage{color}

\usepackage[labelfont=bf,labelsep=period,justification=raggedright]{caption}

\makeatletter
\renewcommand{\@biblabel}[1]{\quad#1.}
\makeatother

\date{}

\begin{document}

% Title must be 150 characters or less
\begin{flushleft}
{\Large
\textbf{Cellular Adaptation Accounts for the Sparse and Reliable
Sensory Stimulus Representation
}}
\\
\vspace{5mm}
Farzad Farkhooi$^{1,\ast}$, Anja Froese$^{2}$, Eilif Muller$^{3}$,
Randolf Menzel$^{2}$, Martin P. Nawrot$^{1}$
\\
\vspace{5mm}
\bf{1} Neuroinformatics \& Theoretical Neuroscience, Freie
Universit\"at Berlin, and Bernstein Center for Computational
Neuroscience Berlin, Berlin, Germany.
\\
\bf{2} Institute f\"ur Biologie-Neurobiologie, Freie Universit\"at
Berlin, Berlin, Germany.
\\
\bf{3} Blue Brain Project, \'Ecole Polytechnique F\'ed\'erale de
Lausanne, Lausanne, Switzerland.
\\
$\ast$ Corresponding author: K\"onigin-Luise-Strasse 1-3, 14195 Berlin, Germany \\
farzad.farkhooi@fu-berlin.de
\end{flushleft}

%\vspace{2cm}

%Abbreviated title: Sequential Adaptation Effects in Sensory Pathways

%Number of pages: 16

%Number of figures: 4

%Number of words for Abstract: 144

%Number of words for Introduction: 348

%Number of words for Discussion: 1033

%Conflict of Interest: The authors declare no competing financial interests.

%\vspace{5mm}

%\newpage

\section*{Abstract}
Most neurons in peripheral sensory pathways initially respond
vigorously when a preferred stimulus is presented, but adapt as
stimulation continues. It is unclear how this phenomenon affects
stimulus representation in the later stages of cortical sensory
processing. Here, we show that a temporally sparse and reliable
stimulus representation develops naturally in a network with adapting
neurons. We find that cellular adaptation plays a critical role in the
transient reduction of the trial-by-trial variability of cortical
spiking, providing an explanation for a wide-spread and hitherto
unexplained phenomenon by a simple mechanism. In insect olfaction,
cellular adaptation is sufficient to explain the emergence of the
temporally sparse and reliable stimulus representation in the mushroom
body, independent of inhibitory mechanisms. Our results reveal a
computational principle that relates neuronal firing rate adaptation
to temporal sparse coding and variability suppression in nervous
systems with a sequential processing architecture.

%\section*{Author Summary}
%Many lines of evidence suggest that few spikes carry the relevant
% stimulus information at later stages of sensory processing.  Yet
% mechanisms for the emergence of sparse sensory representations
% remain unclear.  Here, we introduce an idea in which a temporal
% sparse and reliable stimulus representation develops naturally in
% spiking networks. It combines principles of signal propagation with
% the commonly observed mechanism of neuronal firing rate adaptation.
% Using a stringent mathematical approach we show how a dense rate
% code at the periphery translates into a temporal sparse
% representation in the cortical network. At the same time it
% dynamically suppresses the trial-by-trial variability matching the
% experimental observations in sensory cortices.  Computational
% modeling and experimental measurements suggest that the same
% principle underlies the prominent example of temporal sparse coding
% in the insect mushroom body.  Our results reveal a computational
% principle that relates neuronal firing rate adapation to temporal
% sparse coding and variability suppression in nervous systems.

\section*{Introduction}
The phenomenon of spike-frequency adaptation
(SFA)\cite{adrian_impulses_1926} (also known as spike-rate adaptation)
is a fundamental process in nervous systems, which attenuates the
neuronal stimulus responses to a lower level following an initial high
firing. This process can be mediated by different cell-intrinsic
mechanisms that involve a spike-triggered self-inhibition, and which
can operate in a wide range of time scales\cite{benda_universal_2003,
  lundstrom_fractional_2008}. These processes are probably related to
the early evolution of the excitable
membrane\cite{rudy_diversity_1988,ranganathan_evolutionary_1994} and
are common to vertebrate and invertebrate neurons, both in the
peripheral and central nervous
system\cite{wark_sensory_2007}. Nonetheless, the functional
consequences of SFA in peripheral stages of sensory processing on the
representation of stimuli in later network stages remain unclear.  For
instance, while olfactory stimuli generally display a slow kinetics,
the early olfactory system shows temporally adaptive and precise onset
responses\cite{shusterman_precise_2011,krofczik_rapid_2008}, that may
translate into narrow integration windows for the principal neurons in
the next processing stage, the piriform cortex pyramidal cells in
rodents\cite{poo_odor_2009} and the mushroom body Kenyon cells in
insects\cite{perez-orive_oscillations_2002}. However, it remains
unclear how sparse single-neuron responses can reliably encode
information in face of the response
variability\cite{stein_neuronal_2005} and the sensitivity of cortical
networks to small perturbations\cite{london_sensitivity_2010}.

Here, we show that the SFA mechanism introduces a dynamical
non-linearity in the transfer function of neurons. Subsequently, the
response onset becomes progressively sparser when transmitted across
successive processing stages.  Additionally, the self-regulating
effect of SFA causes a stimulus-triggered reduction of firing
variability by modulating the average inhibition in the balanced
cortical network. In this manner the temporally sparse representation
is accompanied by an increased response reliability. Using this
theoretical framework on data obtained from the olfactory system of
the honeybee, we show that the sequential effect of the SFA shapes the
Kenyon cells' temporal sparseness independent of inhibitory synaptic
mechanisms.

Our results reveal a generic, biophysically plausible mechanism that
can explain the emergence of a reliable and sparse stimulus
representation, indicating the importance of cellular adaptation in
sensory computation at the network level.

\section*{RESULTS}

\subsection*{Temporal sparseness emerges in successive adapting
  populations}

%FIG1
\begin{figure}[!t]
\includegraphics[scale=1.4]{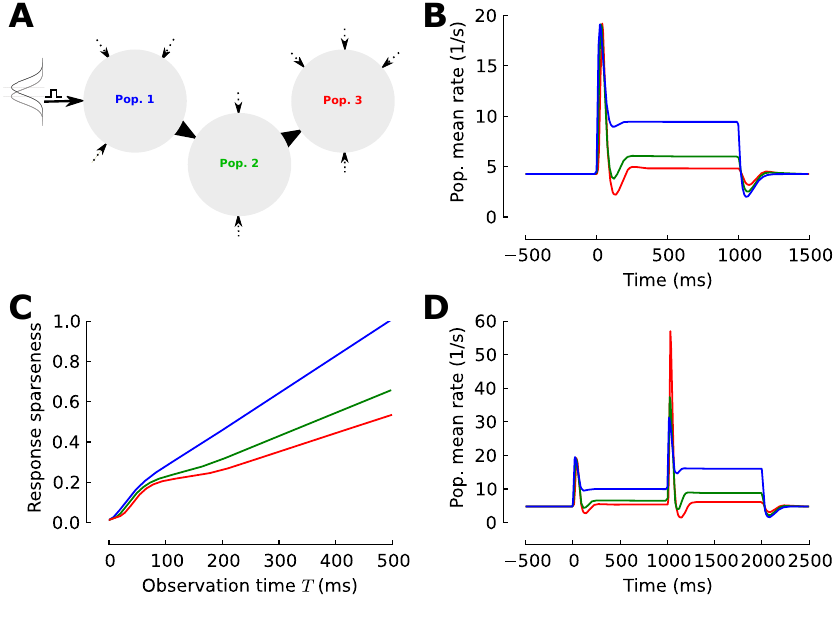}
  \caption{ {\bf Neuronal adaptation in the multi-stage processing network.}
    ({\bf A}) Schematic illustration of a three-layered model of an
    adaptive pathway of sensory processing. The network consists of
    three consecutive adaptive populations. Each population receives
    sensory input from an afferent source (black arrows) and
    independent constant background excitation (dashed arrow). Input
    is modeled by a Gaussian density and a sensory stimulus presented
    to the first population is modeled by an increase in the mean
    input value. ({\bf B}) Response profiles. The evoked state
    consists of a phasic-tonic response in all populations. The tonic
    response level is decremented across the consequtive
    populations. ({\bf C}) Temporal sparseness $S_{r}^i$ is measured
    by the integral over the firing rate and normalized by the average
    spike count at $t=500\,$ms in the first population. Responses
    become progressively sparser as the stimulus propagates into the
    network. ({\bf D}) Secondary response profiles. The additional
    jump increase in stimulus strength at $t=1000\,$ms
    during the evoked state of the first stimulus results in a
    secondary phasic response in all populations with an amplitude
    overshoot in the 2nd and 3rd population.}
\end{figure}

To examine how successive adapting populations can achieve temporal
sparseness, first we mathematically analyzed a sequence of neuronal
ensembles ({\bf Figure 1A}), where each ensemble exhibits a generic
model of mean firing rate adaptation by means of a slow negative
self-feedback\cite{benda_universal_2003, lacamera_minimal_2004,
  muller_spike-frequency_2007} ({\bf Materials and Methods}).  This
sequence of neuronal ensembles should be viewed as caricature for
distinct stages in the pathway of sensory processing. For instance in
the mammalian olfactory system the sensory pathway involves several
stages from the olfactory sensory neurons to the olfactory bulb, the
piriform cortex, and then to higher cortical areas ({\bf Figure 1A}).

The mean firing rate in the steady-state of a single adaptive
population can be obtained by solving the rate consistency equation,
$\bar r_i = f(\mu_i - \tau_{s} q_s \bar r_i|\sigma_i^2)$,
where $\bar r$ is the equilibrium mean firing rate of the population
$i$, $\mu_i$ and $\sigma_i^2$ are mean and variance of the total input
into the population, respectively, $f$ is the response function
(input-output transfer function, or $fI$-curve) of the population mean
activity, $q_s$ is the quantal conductance of the adaptation mechanism
per unit of firing rate, and $\tau_{s}$ is the firing rate relaxation
time constant\cite{benda_universal_2003, lacamera_minimal_2004,
  muller_spike-frequency_2007}.  It is known that any sufficiently
slow modulation ($1/r_{max} < \tau_s$) linearizes the steady-state
solution, $\bar r_i$, due to the self-inhibitory feedback being
proportional to the firing rate\cite{ermentrout_linearization_1998}.

Here, for simplicity, we studied the case where all populations in the
network exhibit the same initial steady-state rate. This is achieved
by adjustment of a constant background input to population $i$ (doted
arrows, {\bf Figure 1A}), resembling the stimulus irrelevant
interactions in the network. All populations are coupled by the same
strength $g$.  First, we calculated the average firing rate dynamics
of the populations' responses following a step increase in the mean
input to the first layer (black arrow, {\bf Figure 1A}). By solving
the dynamics of the mean firing rate and adaptation level
concurrently, we obtained the mean-field approximation of the
populations' firing rates ({\bf Materials and Methods}). The responses
of each population consisted of a phasic and a tonic part, typical for
adapting neurons.  Here, we plotted the mean firing rate of three
consecutive populations ({\bf Figure 1B}). The phasic response to a
step increase in the input is presented across stages. However, the
tonic response becomes increasingly suppressed in the later stages
({\bf Figure 1B}).  This phenomenon is a general feature of successive
adaptive neuronal populations.  To understand why the tonic response
part, after the adaptation was reinstalled, is increasingly suppressed
in later stages we used the linearity of the adaptive transfer
function and its stability conditions at the
steady-state\cite{ermentrout_linearization_1998}. Thus, a change in
the mean rate of population $i$ is the solution of
$\Delta {\bar r}_i = f(g \Delta \bar{r}_{i-1}- \Delta {\bar
  r}_i \tau_s q_s)$,
and since the transfer function, $f$, is assumed to be increasing and
the slope of $\bar r$ is less than unity, the solution can only existe
for $\Delta {\bar r}_{i-1} > \Delta {\bar r}_{i}$ ({\bf Materials and
  Methods}). The result in this sub-section ({\bf Figure 1B-C}) was
established with a current based leaky integrated-and-fire response
function. However, the analysis presented here extends to the majority
of neuronal transfer functions since the stability and linearity of
the adapted steady-states are
granted\cite{ermentrout_linearization_1998,lacamera_minimal_2004}.
This simple effect leads to a progressively sparser representation
across successive stage of a generic feed-forward adaptive processing.
We quantified the temporal sparseness by $S_r^i(T)=\int_0^T r_i(t)dt$,
where $ r_i(t)$ is the mean firing rate of population $i$ and $T$ is
the observation time window. Normalization of this measure by the
first population $S_{r}^1(T = 500)$ indicates sparser responses in
later stages of the adaptive network ({\bf Figure 1C}).

Does the suppression of the adapted response level impair the
information about the presence of the stimulus?  To explore this, we
studied a secondary increase in stimulus strengths of equal magnitude
after 1 second when the network has relaxed to the evoked equilibrium
({\bf Figure 1D}). The secondary stimulus jump induced a secondary
phasic response of comparable magnitude in the first population ({\bf
  Figure 1D}). However, in the later populations this jump evoked an
icreased peak rate in the phasic response ({\bf Figure 1D}). Notably,
the coupling factor $g$ between the populations shapes this
phenomena. Here, we adjusted $g$ to achieve an equal onset response
magnitude across the populations for the first stimulus jump at $t=0$,
and a slight increase in the population onset response in the first
population is amplified in the later stages. This is due to the fact
that the later stages accumulated less adaptation in their evoked
steady-state (level of adaption is proportional to mean firing
rate). Importantly, this result confirms that the sustained presence
of the stimulus is indeed stored in the level of cellular
adaptation\cite{nesse_biophysical_2010}, even though it is not
reflected in the firing rate of the last population.
Therefore, regardless of the absolute amplitude of responses, the
relative relation between secondary and initial onset keeps increasing
across layers. This type of a secondary overshoot is also
experimentally known as {\it sensory sensitization}, where an
additional increase in the stimulus strength significantly enhances
the responsiveness of later stages after the network converged to an
adapted steady-state\cite{kadohisa_olfactory_2006}.

\subsection*{Adaptation increases response reliability in the
  cortical network}

%FIG2
\begin{figure}[t!]
\includegraphics[scale=0.9]{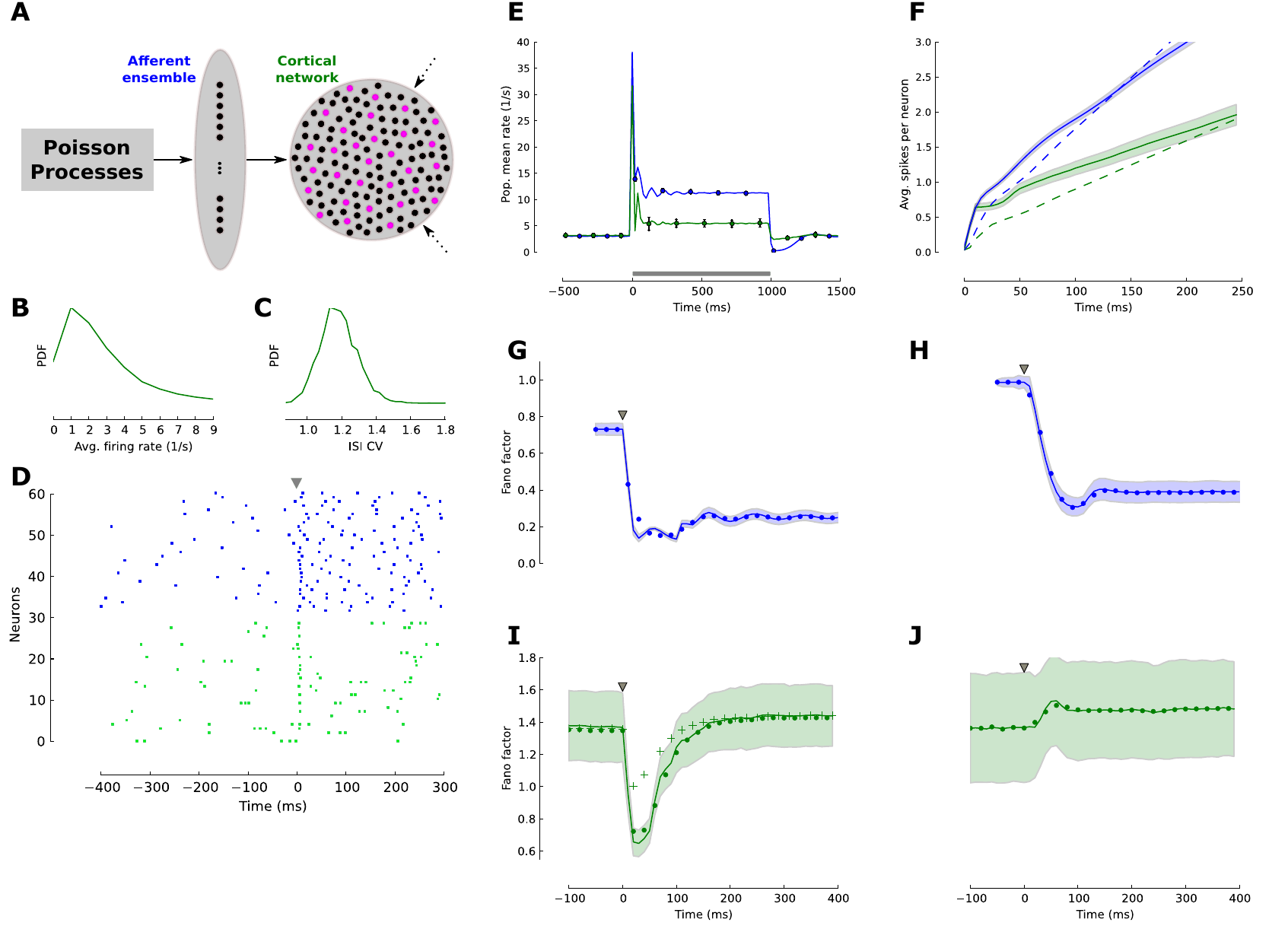}
  \caption{{\bf Reliability of a temporally sparse code in the balanced
    cortical network.}  ({\bf A}) Schematic of a two-layer model of
    sub-cortical and early cortical sensory processing. The afferent
    ensemble (blue) consists of 4,000 independent neurons, and each
    neuron projects to 1\% of the neurons in the cortical network
    (green). The cortical network is a balanced network in the
    asynchronous and irregular state with random connectivity. In both
    populations black circles represent excitatory neurons and magenta
    circles represent inhibitory cells. ({\bf B}) The distribution of
    firing rates across neurons in the cortical network is fat-tailed
    and the average firing rate is approximately 3Hz. ({\bf C}) The
    distribution of the coefficient of variation ($C_v$) across neurons
    in the balanced cortical network confirms irregular spiking. ({\bf
      D}) Spike raster plot for a sample set of 30 afferent neurons
    (blue dots) and 30 excitatory cortical neurons (green dots). At
    $t=0$ (gray triangle) the stimulus presentation starts. ({\bf E})
    Population averaged firing for both network stages. The simulation
    (solid lines) follows the calculated ensemble average predicted by
    the adaptive density treatment (black circles). The firing rate in
    the simulated network is estimated with a 20ms bin size. ({\bf F})
    Number of spikes per neuron $S_e$ after stimulus onset ($t=0$) for
    the adaptive network (solid lines) and the weakly adaptive control
    network (dashed lines). Cortical excitatory neurons (green)
    produced less spikes than neurons in the earlier stage (afferent
    ensemble, blue).  The shaded area indicates the standard deviation
    across neurons.  ({\bf G,H}) Fano factor dynamics of the afferent
    ensemble in the network with strongly adapting neurons (G,
    $\tau_s=110\,$ms) and in the weakly-adaptive (h, $\tau_s=30\,$ms)
    network, estimated across 200 trials in a 50ms window and a
    sliding of 10ms for the ensemble network with adaptation. The
    black circles indicate the theoretical value of the Fano factor
    computed by adaptive density treatment and shaded area is the
    standard deviation of the Fano factor across neurons in the
    network. ({\bf I}) The Fano factor of strongly adaptive neurons in
    cortical balanced network reduced transiently during the initial
    phasic response part. The crosses show the adaptive cortical
    ensemble Fano factor for the case where the afferent ensemble
    neurons were modeled as a Poisson process with the same
    steady-state firing rate and without adaption. ({\bf J}) The Fano
    factor in the weakly adaptive cortical network did not exhibit a
    reduction during stimulation.}
\end{figure}

The mean firing rate approach as above is insufficient to determine
how reliable the observed response transients are across repeated
stimulations. In the prevailing spiking network model of the cortex,
the balance of excitation and inhibition is quickly reinstalled within
milliseconds after the onset of an excitatory
input\cite{van_vreeswijk_chaotic_1998} and self-generating recurrent
fluctuations strongly dominate the dynamics of the interactions. It
has thus been questioned whether a sparse response of only very few
action potentials could reliably encode the presence of a
stimulus\cite{london_sensitivity_2010}.

To investigate the reliability of adaptive mapping from a dense
stimulus to a sparse cortical spkie response across successive
processing stages we employed an adaptive population density
formalisms\cite{muller_spike-frequency_2007, farkhooi_adaptation_2011}
({\bf Materials and Methods}) along with numerical network
simulations.
We embeded a two-layered sensory network with an afferent ensemble
projecting to a cortical network ({\bf Figure 2A}). The afferent
ensemble consisted of 4,000 adaptive neurons that included voltage
dynamics, conductance-based synapses, and spike-induced
adaptation\cite{muller_spike-frequency_2007}.  It resembles the
sub-cortical sensory processing and each neuron in the afferent
ensemble projects to randomly 1\% of the neurons in the cortical
network.  This is a large circuit of the balanced network ({\bf Figure
  2A}) with 10,000 excitatory and 2,500 inhibitory neurons with a
typical random diluted connectivity of 1\%. The spiking neuron model
in the cortical network again includes voltage dynamics,
conductance-based synapses, and spike-induced
adaptation\cite{muller_spike-frequency_2007}. All neurons are alike
and parameters are given in {\bf Table.3} in Muller et
al.\cite{muller_spike-frequency_2007}.  With appropriate adjustment of
the synaptic weights, the cortical network operates in a globally
balanced manner, producing irregular, asynchronous
activity\cite{van_vreeswijk_chaotic_1998,
  kumar_high-conductance_2008,vogels_gating_2009}. The distribution of
firing rates for the network approximates a power-law
density\cite{roxin_distribution_2011} with an average firing rate of
$\approx 3.0\,$Hz ({\bf Figure 2B}) and the coefficients of variation
($C_v$) for the inter-spike intervals is centered at a value slightly
greater than unity ({\bf Figure 2C}) indicating the globally balanced
and irregular state of the network\cite{vogels_gating_2009}.
Noteworthy, the activity of neurons in both stages is fairly
incoherent and spiking in each sub-network is independent. Therefore,
one can apply an adiabatic elimination of the fast variables and
formulate a population density description where a detailed neuron
model reduces to a stochastic point process
\cite{muller_spike-frequency_2007,farkhooi_adaptation_2011} that
provides an analytical approximation and understanding of the
simulation results in this section ({\bf Materials and Methods}).

The background input is modeled as a set of independent Poisson
processes that drive both sub-networks (dashed arrows, {\bf
 Figure 2A}). The stimulus dependent input is an increase in the
intensity of the Poisson input into the afferent ensemble (solid
arrow, {\bf Figure 2A}).   Before  the  stimulus became active at time
$t=0$, a  typical  neuron  showed  an  irregular  spiking activity in
both network stages ({\bf Figure 2B-D}). However,  when  a  sufficiently 
strong  stimulus  is  applied, neurons in both stages exhibited a
transient response before the population mean firing rate converges
back to a new level of steady-state ({\bf Figure 2D,E}).  The population
firing rate in the numerical simulations (solid line, {\bf Figure 2E})
follow well the adaptive population density treatment (filled circles,
{\bf Figure 2E}).

To measure the effect of neuronal adaptation on the temporal
sparseness, we again computed the number of spikes per neuron after
the stimulus onset, $S_e(T)$. We compare our standard adaptive network
with an adaptation time constant of $\tau_{s}=$ 110ms (solid lines,
{\bf Figure 2F}) with an only weakly-adaptive control network
($\tau_{s}=$30ms; dashed lines, {\bf Figure 2F}). Note, that the
adaptation time constant in the weakly adaptive network is about equal
to the membrane time constant and therefore plays a minor role for the
network dynamics.  It showed that both sub-networks generated sharp
phasic response, which in the case of the cortical network evoked a
single sharply timed spike within the first $\approx 10-20\,$ms in a
subset of neurons ({\bf Figure 2B,F}). In the control case, the
response is non-sparse and response spikes are distrubuted throughout
the stimulus period (dashed lines, {\bf Figure 2F}). Overall, strong
adaptation reduces the total number of stimulus-induced action
potentials per neuron and concentrates their occurrence within an
initial brief phase following fast changes in the stimulus.  Thus, in
accordance with the results of the rate-based model in the previous
section, one can conclude that the sequential stimulation of adaptive
network stages can account for the emergence of a sparse stimulus
representation in a cortical population.

To  reveal the effect of adaptation on the response variability, we
employed the time-resolved Fano factor\cite{nawrot_measurement_2008},
$F$, which measures the spike-count variance divided by the mean spike
count across $200$ repeated stimulations. Spikes were counted in a
50ms time window and a sliding of
10ms\cite{churchland_stimulus_2010}. As before, we compared our
standard adaptive network ({\bf Figure 2G,I}; $\tau_{s}$ = 110ms) with
the control network ({\bf Figure 2H,J}; $\tau_{s}$=30ms). Since the
Fano factor is known to be strongly dependent on the firing rate, we
adjusted the input to the latter such that the steady state firing
rates in both networks were
mean-matched\cite{churchland_stimulus_2010}.  The input Poisson spike
trains ($F=1$) translated into slightly more regular spontaneous
($t<0$) activity in the afferent ensemble ({\bf Figure 2G}), as neuronal
membrane filtering and refractoriness reduced the output
variability. After the stimulus onset ($t=0$), due to the increase in
the mean input rate, the average firing rate increased, however the
variance of the number of events per trial did not increase
proportionally. Therefore, we observed a reduction in the Fano factor
({\bf Figure 2G}). This phenomena is independent of the adaptation
mechanism in the neuron model and a quantitatively similar reduction
can be observed in the weakly adaptive afferent ensemble ({\bf Figure
  2H}). A comparison between our standard adaptive and the control
case reveals that the adaptive network is generally more regular in
the background and in the evoked state ({\bf Figure 2G,H}). This is
due to the previously known effect, where adaptation induces negative
serial correlations in the inter-spike intervals\cite{
  benda_linear_2010,farkhooi_adaptation_2011} and as a result reduces
the Fano factor\cite{farkhooi_adaptation_2011,
  chacron_electroreceptor_2005}.

In the next stage of processing, the distribution of $F$ across
neurons during spontaneous activity is high due to the self-generated
noise of the balanced circuits\cite{van_vreeswijk_chaotic_1998,
  lerchner_response_2006}.  This closely follows a wide spread
experimental finding where $F > 1$ in the spontaneous cortical
activity\cite{churchland_stimulus_2010} ( {\bf Figure 2G,H}). Whenever
a sufficiently  strong  stimulus  was applied, the  internally
 generated  fluctuations  in the adaptive balanced network were
transiently suppressed,  and as a result the Fano factor dropped
sharply ({\bf Figure 2I}). However, this reduction of the Fano factor
is a temporary phenomenon and $F$ converges back to slightly above the
baseline variability ({\bf Figure 2I}). At the same time, the evoked
steady-state firing returned back to the irregular and asynchronous
state ({\bf Figure 2D}).  Indeed, this effect corresponds to a
temporally mis-match in the balanced input conditions to the cortical
neurons since the self-inhibitory and slower adaption effect prevents
a rapid adjustment to the new input regime. This can be observed in
the time course of variability suppression that closely reflects the
time constant of adaptation ({\bf Figure 2I}). The afferent ensemble
structured the input to the cortical ensemble, determining the
magnitude of the observed reduction effect, and under the control
condition where a pure Poisson input is provided to the cortical
balanced network, the reduction in $F$ is reduced but the time scale
of recovery remains unaltered (crosses, {\bf Figure 2I}). We  contrast
 this adaptive  behavior  with the variability dynamics in the weakly
adaptive balanced network ({\bf Figure 2J}). In  this  case there  is
 no  reduction  in  $F$, because for a short adaptation time constant
the convergence to the balanced state is very rapid\cite{
  van_vreeswijk_chaotic_1998}. The small increase in the input noise
strength leads to an increase of the self-generated randomness of the
balanced
network\cite{monteforte_dynamical_2010,lerchner_response_2006}.

To further understand the effect of the recurrent connectivity on the
variability, one can compare the evoked variability between the
afferent ensemble and the cortical network.  In the latter,
fluctuation intrinsic to the cortical networks dynamically bring back
the cortical circuits to the spontaneous level of high variability
within the time scale of adaption relaxation ({\bf Figure 2I}).  This
suggests the transient suppression of variability during a stimulus
response as a cortical-wide effect\cite{churchland_stimulus_2010}
could be caused by slow adjustment of self-regulatory feedback.

%\newpage

\subsection*{Adaptive networks generate sparse and reliable responses in the insect olfactory system}

%FIG3
\begin{figure}[!t]
\includegraphics[scale=0.9]{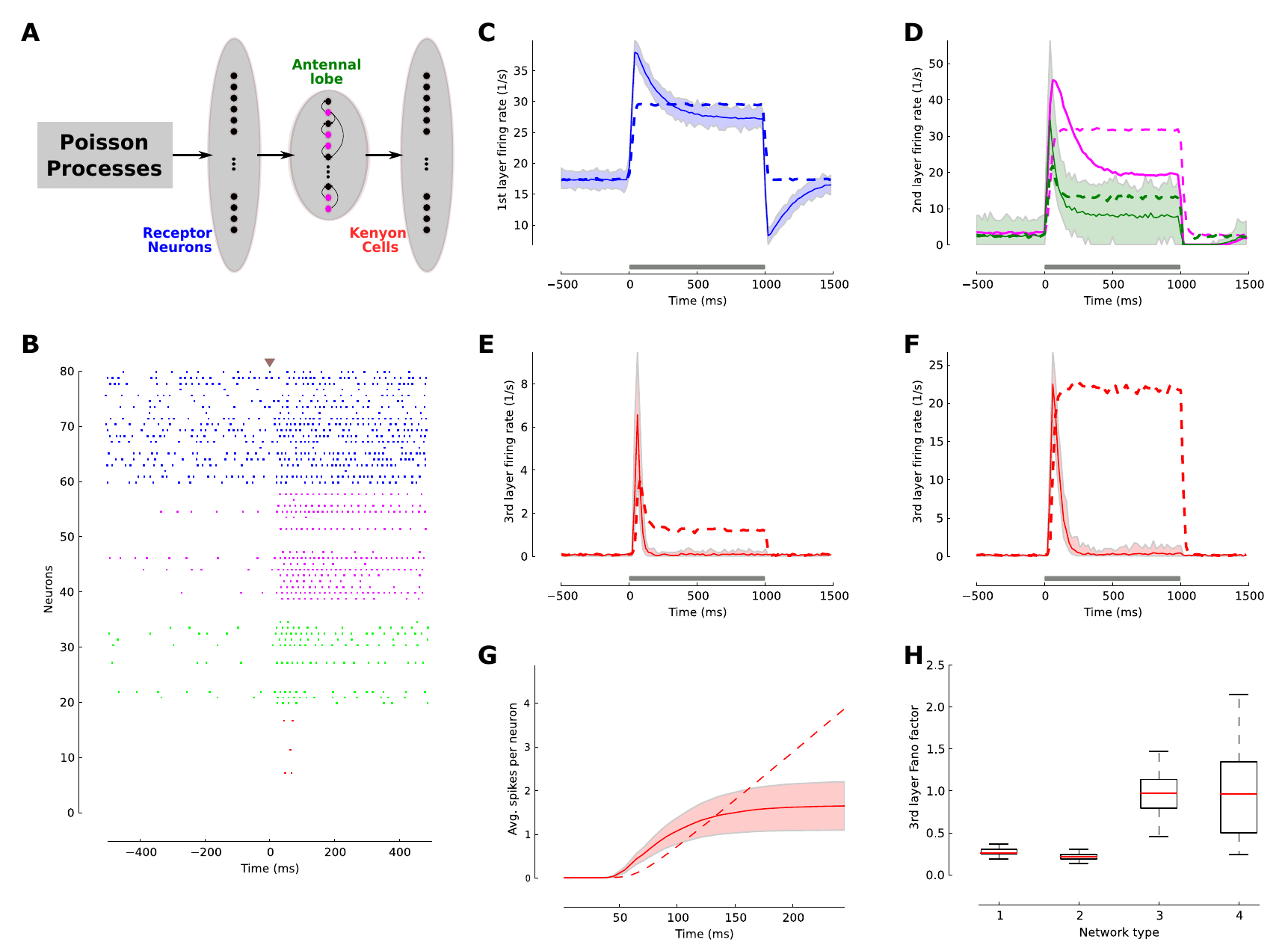}
  \caption{{\bf Neuronal adaptation generates temporal sparseness in a
    generic model of the insect olfactory network.}  ({\bf A})
    Schematic drawing of a simplified model of the insect olfactory
    network for a single pathway of odor coding. Olfactory receptor
    neurons (OSNs, first layer, n=1,480) project to the antennal lobe
    network (second layer) consisting of projection neurons (PNs,
    n=24) and local neurons (magenta, n=96), which make inhibitory
    connections with PNs. PNs project to the Kenyon cells (KCs) in the
    mushroom body (third layer). ({\bf B}) Spike raster plot of
    randomly selected OSNs (blue), LNs (magenta), PNs (green) and KCs
    (red) indicates that spiking activity in the network became
    progressively sparser as the Poisson input propagated into the
    network. ({\bf C}) Average population rate of OSNs in the adaptive
    network (blue solid line) and the non-adaptive control network
    (dashed blue lines). The shaded area indicates the firing rate
    distribution of the neurons. The firing rate was estimated with
    20ms bin size. ({\bf D}) Average response in the antennal lobe
    network.  PNs (green) and LNs (magenta) exhibited the typical
    phasic-tonic response profile in the adaptive network (solid
    lines) but not in the non-adaptive case (dashed lines).  ({\bf E})
    Kenyon cell activity. In the adaptive network the KC population
    exhibits a brief response immideately after stimulus onset, which
    quickly returns close to baseline. This is contrasted by a tonic
    response profile throughout the stimulus in the non-adaptive case.
    ({\bf F}) Effect of the inhibitory micro-circuit. By turning off
    the inhibitory LN-PN connections the population response amplitude
    of the KCs was increased, while the population response dynamics
    did not change.({\bf G}) Sparseness of KCs. The average number of
    spikes per neuron emitted since stimulus onset indicates that the
    adaptive ensemble encodes stimulus informatoin with only very few
    spikes. ({\bf G}) Reliability of KCs responses. The Fano factor of
    the KCs in different network scenarios is estimated across 200
    trials in a 100ms time window after stimulus onset. Network 1:
    (+)Adaptation (+)Inhibition, network 2: (+)Adaptation
    (-)Inhibition, network 3: (-)Adaptation (-)Inhibition, network 4:
    (-)Adaptation (+)Inhibition. Both networks with SFA are
    significantly more reliable in their stimulus encoding than the
    non-adaptive networks.}
\end{figure}

As a case study, we investigated the role of adaptation for the
emergence of the sparse temporal code in the insects olfactory system,
which is analogoue to the mammalian olfactory system.  We simulated a
reduced generic model of olfactory processing in insects using the
adaptive neuron model\cite{muller_spike-frequency_2007}.  The model
network consisted of an input layer with 1,480 olfactory sensory
neurons (OSNs), which project to the next layer representing the
antennal lobe circuit with 24 projection neurons (PNs) and 96
inhibitory local inter-neurons (LNs) that form a local feed-forward
inhibitory micro-circuit with the PNs. The third layer holds 1,000
Kenyon cells (KCs) receiving divergent-convergent input from PNs. The
relative numbers for all neurons approximate the anatomical ratios
found in the olfactory pathway of the
honeybee\cite{menzel_olfaction_2009} ({\bf Figure 3A}). We introduced
heterogeneity among neurons by randomizing their synaptic time
constants and the connectivity probabilities are chosen according to
anatomical studies. Synaptic weights were adjusted to achieve
spontaneous firing statistics that match the observed physiological
regimes. The SFA parameters were identical throughout the network with
$\tau_{s}=110\,$ms (see {\bf Materials and Methods} for details).

Using this model, we sought to understand how adaptation contributes
to temporally sparse odor representations in the KC layer in a small
sized network and under highly fluctuating input conditions. We
simulated the input to each OSN by an independent Poisson process,
which is thought to be reminiscent of the transduction process at the
olfactory receptor level\cite{nagel_biophysical_2011}. Stimulus
activation was modeled by a step increase in the Poisson intensity
with uniformly jittered onset across the OSN population ({\bf
  Materials and Methods}).  Following a transient onset response the
OSNs adapted their firing to a new steady-state ({\bf Figure
  3B,C}). The pronounced effect of adaptation becomes apparent when
the adaptive population response is compared to the OSN responses in
the control network without any adaptation ($\tau_s$=0; dashed line,
{\bf Figure 3C}). In the next layer, the PN population activity is
reflected in a dominant phasic-tonic response profile (green line,
{\bf Figure 3D}), which closely matches the experimental
observation\cite{krofczik_rapid_2008}. This is due to the
self-inhibitory effect of the SFA mechanism, and to the feedforward
inhibition received from the LNs (magenta line, {\bf Figure
  3D}). Consequently, the KCs in the third layer produced only very
few action potentials following the response onset (red line , {\bf
  Figure 3E,G}).  The average number of emitted KC response spikes per
neuron, $S_e$, is small in the adaptive network (average $<2$) whereas
KCs continue spiking throughout stimulus presentation in the
non-adaptive network ({\bf Figure 3G}). This finding closely resembles
experimental findings of temporal sparseness of KC responses in
different insect species\cite{perez-orive_oscillations_2002,
  broome_encoding_2006, szyszka_sparsening_2005} and quantitatively
matches the KC response statistics provided by Ito and
colleagues\cite{ito_sparse_2008}.  The simulation results obtained
here confirm the mathematically derived results in the first results
section ({\bf Figure 1}) and show that neuronal adaption can cause a
temporal sparse representation even in a fairly small and highly
structured layered network where the mathematical assumptions of
infinite network size and fundamentally incoherent activity are not
full filled.

To test the effect of inhibition in the LN-PN micro-circuitry within
the antennal lobe layer on the emergence of temporal sparseness in the
KC layer, we deactivated all LN-PN feedforward connections and kept
all other parameters fix. We found a profound increase in the
amplitude of the KC population response, both in the adaptive (red
line,{\bf Figure 3F}) and the non-adaptive network (dashed red
line,{\bf Figure 3F}). This increase in response amplitude is carried
by an increase in the number of responding KCs due to the increased
excitatory input from the PNs, implying a strong reduction in the KC
population sparseness. Importantly, removing local inhibition did not
alter the temporal profile of the KC population response in the
adaptive network (cf. red lines in {\bf Figure 3E,F}), and thus
temporal sparseness was independent of inhibition in our network
model.

How reliable is the sparse spike response across trials in single KCs?
To answer this question, we again measured the robustness of the
stimulus representation by estimating the Fano factor across $200$
simulation trials ({\bf Figure 3H}). The network with adaptive neurons
and inhibitory micro-circuitry exhibited a low Fano factor (median
$\approx 0.3$) and a narrow distribution across all neurons. This
follows the experimental finding that the few spikes emitted by KCs
are highly reliably\cite{ito_sparse_2008} (network 1, {\bf Figure
  3H}). Turning off the inhibitory micro-circuitry did not
significantly change the response reliability (Wilcoxon rank sum test,
p-value = 0.01; network 2, {\bf Figure 3H}). However, both networks
that lacked adaptation exhibited a significantly increased variability
with a median Fano factor close to one (Wilcoxon rank sum test,
p-value = 0.01; networks 3 and 4, {\bf Figure 3H}), independent of the
presence or absence of inhibition within the antennal lobe.

% FIG4
\begin{figure}[!t]
\includegraphics[scale=1.4]{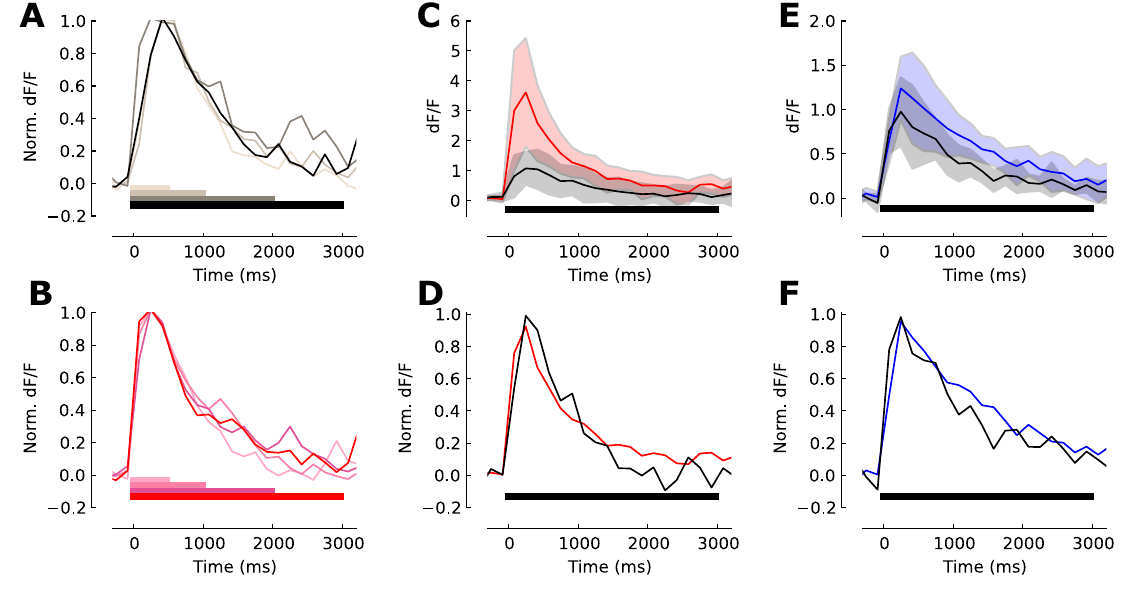}
  \caption{ {\bf Blocking GABAergic transmission in the honeybee changes
    amplitude but not duration of the KC population response.} ({\bf
      A}) Temporal response profile of the Ca signal imaged in the
    mushroom body lip region of one honeybee for different stimulus
    durations as indicated by color. ({\bf B}) Temporal response
    profiles as in {(\bf A)} in one honeybee after application of
    PTX. ({\bf C}) Response profiles imaged from 6 control animals
    (gray) and their average (black) for a 3s stimulus as indicated by
    the stimulus bar. The responses measured in 6 animals in which
    GABA$_A$ transmission was blocked with PTX (red) shows a
    considerably higher population response amplitude. The shaded area
    indicated the standard deviation of responses across bees.  ({\bf
      D}) Average amplitude-normalized responses are highly similar in
    animals treated with PTX and control animals. ({\bf E}) Blocking
    GABA$_B$ transmission with CGP in 6 animals (blue) again results
    in an increased response amplitude compared to 6 control animals
    (black). The shaded area indicates the standard deviation across
    individuals. ({\bf F}) Average normalized response profiles are
    highly similar in the CGP-treated and control animals.}
\end{figure}

To explore whether neuronal adaptation could be the responsible
mechanism for temporal sparseness in the biological network, we
performed a set of Calcium imaging experiments, monitoring Calcium
responses in the KC population of the honeybee mushroom
body\cite{szyszka_sparsening_2005} ({\bf Materials and Methods}). Our
computational model ({\bf Figure 3}) predicted that blocking of the
inhibitory microcircuit would increase the population response
amplitude but should not alter the temporal dynamics of the KC
population response.  In a set of experiments, we tested this
hypothesis by comparison of the KCs' evoked activity in the presence
and absence of GABAergic inhibition ({\bf Materials and Methods}).
First, we analyzed the normalized Calcium response signal within the
mushroom body lip region in response to a 3s, 2s, 1s and 0.5s odor
stimulus ({\bf Figure 4A}).  We observed the same brief phasic
response following stimulus onset in all four cases with a
characteristic slope of Calcium response decay that has been reported
previously\cite{szyszka_sparsening_2005}.  Bath application of the
GABA$_A$ antagonist picrotoxin (PTX) did not change the time course of
the Calcium response dynamics ({\bf Figure 4B-D}).  The effectiveness
of the drug was verified by the increased population response
amplitude in initial phase ({\bf Figure 4C}). Next, we tested the
GABA$_B$ antagonist hydrochloride (CGP) using the same protocol and
again found an increase in the response magnitude but no alteration of
the response dynamics ({\bf Figure 4E,F}).

\section*{DISCUSSION}

We here propose that a simple neuron-intrinsic mechanism of
spike-triggered adaptation can account for a reliable and temporally
sparse sensory stimulus representation across stages of sensory
processing. At the single neuron level SFA is known to induce the
functional property of a fractional differentiation with respect to
the temporal profile of the input and thus offers the possibility of
tuning the neuron's response properties to the relevant stimulus
time-scales at the cellular level\cite{benda_spike-frequency_2005,
  benda_universal_2003, lundstrom_fractional_2008}. Our results
indicate that a sensory processing in a feedforeward network with
adaptive neurons focuses on the temporal changes of the sensory input
in a precise and temporally sparse manner ({\bf Figure 1B; Figure 2E
  and Figure 3E}). At the same time the constancy of the stimulus is
memorized in the cellular level of
adaptation\cite{nesse_biophysical_2010} ({\bf Figure 1D}). We hypothesize
that at later processing stages in the sensory pathway, notably in the
sensory cortices and in the insect mushroom body, it is most relevant
to encode and process relevant dynamic \emph{changes} in the sensory
environment and to neglect the static stimulus features as well as the
dynamic fluctuations of receptor sensations present at very fast time
scales.  One prominent example is primate vision where, in the absence
of the self-generated dynamics of retina input due to microsaccades,
observers become functionally blind to stationary objects during
fixations\cite{martinez-conde_microsaccades:_2009}.

The emergence of a sparse representation has been demonstrated in
various cortical sensory areas, for example in
visual\cite{tolhurst_sparseness_2009},
auditory\cite{hromadka_sparse_2008},
somatosensory\cite{jadhav_sparse_2009}, and
olfactory\cite{poo_odor_2009} cortices, and thus manifests a principle
of sensory computation across sensory modalities and independent of
the natural stimulus kinetic. However, it has been repeatedly
questioned whether a few informative spikes can survive in the
cortical network, which is highly sensitive to small
perturbations\cite{wolfe_sparse_2010,london_sensitivity_2010}. Our
results show that a biologically realistic cellular mechanism
implemented at successive network stages can transform a dense and
highly variable Poisson input at the periphery into a sparse and
highly reliable ensemble representation in the cortical network. These
results reflect previous theoretical evidence that SFA has an
extensive synchronizing-desynchronizing effect on population responses
in a balanced network\cite{van_vreeswijk_analysis_2000}.  However,
cellular adaptation has largely been neglected in theory and
simulation of cortical networks, with very few
exceptions\cite{van_vreeswijk_analysis_2000}, although it facilitates
a transition from a dense rate code to a temporal code expressed in
the concerted spiking of cortical cell
assemblies\cite{kumar_spiking_2010} ({\bf Figure 2D}).

Importantly, the adaption level adjusts with a dynamics that is slow
compared to the dynamics of excitatoary and inhibitory synaptic
inputs. This circumstance allows for a transient mismatch of the
balanced state in the cortical network and thus leads to a transient
reduction of the self-generated (recurrent) noise ({\bf
 Figure 2I}). This, in turn, explains why the temporally sparse
representation can be highly reliable. This result seems
counter-intuitive in the light of previous studies that consistently
stressed the persistence of noise in the balanced network\cite{
  lerchner_response_2006, monteforte_dynamical_2010,
  london_sensitivity_2010}. However, cellular adaptation has largely
been neglected in theory and simulation of cortical networks with very
few exceptions\cite{van_vreeswijk_analysis_2000}.

Our results provide an understanding of the role of slow cellular
adaptation dynamics for the transient suppression of the
trial-by-trial response variability. We thus arrived at a coherent
model explanation for the previously unexplained and wide-spread
experimental phenomenon of a stimulus-driven transient reduction of
response variability in cortical neurons, which has been observed in
different species and across many cortical areas, from occipital to
frontal, as e.g. reported by Churchland and colleagues for 14
independent datasets\cite{churchland_stimulus_2010}.

The insect olfactory system is experimentally well investigated and
examplifies a pronounced sparse coding scheme at the level of the
mushroom body KCs. The olfactory system is analog in invertebrates and
vertebrates and the sparse stimulus representation is likewise
observed in the pyramidal cells of the piriform
cortex\cite{poo_odor_2009}, and the rapid responses in the mitral
cells in the olfactory bulb\cite{shusterman_precise_2011} compare to
those of projection neurons in antennal
lobe\cite{krofczik_rapid_2008}.  Our adaptive network model, designed
in coarse analogy to the insect olfactory system, produced
increasingly phasic population responses as the stimulus-driven
acitivty propagated through the network. Our model results closely
match the repeated experimental observation of temporally sparse KC
responses in extracellular recordings from the
locust\cite{perez-orive_oscillations_2002} and
manduca\cite{broome_encoding_2006,ito_sparse_2008}, and in Calcium
imaging in the honeybee\cite{szyszka_sparsening_2005}.

In our experiments we could show that systemic blocking of GABAergic
transmission did not affect the temporal sparseness of the KC
population response in the honeybee ({\bf Figure 4}). This result might
seem to contradict former studies that stressed the role of inhibitory
feed-forward\cite{assisi_adaptive_2007} or feedback
inhibition\cite{papadopoulou_normalization_2011,gupta_functional_2012}
for the emergence of KC sparseness. However, the suggested inhibitory
mechanisms and the sequential effect in the adaptive network proposed
here are not mutually exclusive and may act in concert to establish
and maintain a temporal and spatial sparse code in a rich and dynamic
natural olfactory scene.

The adaptive network model manifests a low trial-to-trial variability
of the sparse KC responses that typically consist of only 1-2
spikes. In consequence, a sparsly activated KC ensemble is able to
robustly encode stimulus information. The low variability at the
single cell level ({\bf Figure 3H}) carries over to a low variability of
the population response\cite{farkhooi_adaptation_2011}. This benefits
downstream processing in the mushroom body output neurons that
integrate converging input from many KCs\cite{menzel_olfaction_2009},
and which were shown to reliably encode odor-reward associations in
the honeybee\cite{strube-bloss_mushroom_2011}.

Our results here bear consequences of general importance for sensory
coding theories. A mechanism of self-inhibition at the cellular level
can facilitate a temporally sparse ensemble code but does not require
a well adjusted interplay between the excitatory and inhibitory
circuitry at the network level. This network effect is robust due to
the distributed nature of the underlying mechanism, which acts
independently in each single neuron. The regularizing effect of
self-inhibition increases the signal-to-noise ratio not only of single
neuron responses but also of the neuronal population
activity\cite{farkhooi_adaptation_2011} that is postsynaptically
integrated in downstream neurons.

\section*{Materials and Methods}

\subsection*{Rate model of generic feedforward adaptive network}

To address analytically the sequential effect of adaptation in a
feedforward network, we consider a model in which populations are
described by their firing rates. Although firing rate models typically
provide a fairly accurate description of network behavior when the
neurons are firing asynchronously\cite{treves_mean-field_1993}, they
do not capture all features of realistic networks. Therefore, we
verify all of our predictions with a population density
formalism\cite{farkhooi_adaptation_2011} as well as a large-scale
simulation of realistic spiking neurons.  To determine the mean
activity dynamics of a consecutive populations, we employed an
standard mean firing rate model of population $i$ as
\begin{equation}
  \begin{cases} \dot{r_i} = - r_i + f(g r_{i-1} +
 \mu_i -a_i|\sigma_i) 
    \\ \dot{a_i} = -\frac{1}{\tau_s} a_i +  r_i q_s  \end{cases}
\end{equation}
where $f$ is the transfer input-output function, $\tau_s$ is the
adaptation time-scale, $g$ is the coupling factor between two
populations and $a$ is the adaptive negative feedback for the
population $i$ with $q_s$ strength and $\sigma$ is the standard
deviation of the input. For simplicity we define $m_i=g r_{i-1} +
\mu_i$. Given $f$ the equilibrium can be determined by
\begin{equation}
  \bar r_i = f(m_i - \tau_s q_s\bar r_i| \sigma_i).
\label{meanRate}\end{equation}
This fix point is always stable\cite{lacamera_minimal_2004} and the
condition for the stability reads $\partial_{r_i} f(m - \tau_sq_s \bar
r_i|\sigma_i) |_{\bar{r}_i} < 1$ and $\partial_{m_i} f(m - \tau_sq_s \bar
r_i|\sigma_i) |_{\bar r} < 0$. It is also known that the $\bar r_i$ is a
linear function, given a sufficiently slow adaptation $\tau_s$ and a
non-linear shape of $f$ \cite{ermentrout_linearization_1998}. Thus, if
$r_{i-1} = r_{i}$ in two subsequent populations, and increased input
to population $i-1$ leads to smaller increase for next population $i$
and the adapted level of responses satisfy $ \Delta r_{i-1} > \Delta
r_{i}$.

The stimilus onset response lies between the adapted steady-state and
not adapted responses function $f$ given the level of new input and
can be analytical calculated accordingly\cite{benda_universal_2003}.

\subsection*{Population density approach to the adaptive neuronal
  ensemble}

In Muller et al.\cite{muller_spike-frequency_2007} it is shown that by
an adiabatic elimination of fast variables a detailed neuron model
including voltage dynamics, conductance-based synapses, and
spike-induced adaptation reduces to a stochastic point process. Thus,
we define an \textit{orderly point process} with a hazard function
argument with state variable $x$ as
\begin{equation}
  h_x(x, t) := \lim_{\Delta t \rightarrow 0^{+}} \frac{ \Pr(N[t , t +  \Delta t ) > 0 | x, t)} {\Delta t }.
\end{equation}
where $N[t , t + \Delta t )$ is the number of events in $\Delta
t$. We assume the dynamic of the adaptation variable is
\begin{equation}
\dot{x} =  - \frac{1}{\tau_s}x + \sum_{k} \delta( t - t _k) q_s,
\end{equation}
where $t_k$ is the time of $k$th spike in the ensemble.
Thus, the state variable distribution at time $t$ in the ensemble is
governed by a master equation of the form
\begin{eqnarray}
  \partial_t \Pr(x, t) &=&  -\partial_{x}\left[ \frac{x}{\tau_s} \Pr(x, t) \right] \nonumber + h_x(x-q_s, t) \Pr(x-q_s, t) \nonumber \\ 
  &-& h_x(x,  t) \Pr(x, t). 
\label{masterEQ}
\end{eqnarray}
We solve the Eq.~\ref{masterEQ} with the help of the transformations
$t_s = \eta(x):= - \tau \log(x/q)$ and $\psi(t_s) = \eta(\eta^{-1}(t_s
+ q))$ numerically\cite{muller_spike-frequency_2007}.
It turns out that indeed $h_x$ is the input-out transfer function of
neurons in the network where its instantaneous parameters are give by
the input statistics\cite{muller_spike-frequency_2007}. Here, we used
the mean-field formalism developed by Lechner et
al.\cite{lerchner_response_2006} to approximately determine the
averaged input within a standard balanced network, as the parameters
of the hazard function $h_x$. Therefore, the functional form of its
solution in a compact form is
\begin{equation}
{\Pr}(t_s|t) = k_t  \Omega(t_s|t)
\label{sol}\end{equation}
where $t_s^0$ is the initial condition state of the system and $k_t$
is a constant defined by $\int {\Pr}(t_s|t)dt_s =1$. Hence, it can be
shown that the firing rate and the consistency equation of the
ensemble is
\begin{equation}
r_t = \int_{-\infty}^{\infty} h_{t_s}(t_s|t) {\Pr}(t_s|t) dt_s.
\label{alpha} \end{equation}
Now, by applying the techniques in Farkhooi et
al\cite{farkhooi_adaptation_2011}, we define a joint probability
density as
 \begin{equation}
 \rho_n ( t_n, t_x^n | t_x^0, t)
 \end{equation}
 where an $n^\mathrm{th}$ event occurs at time $t_n > t$ and the state
 variable is $t_x^n$. We can write $n+1^\mathrm{th}$ event time and
 state of adaptation joint density recursively
\begin{equation}
  \rho_{n+1}(t_{n+1} , t_x^{n+1} | t_x^0, t)= \int\int \rho( t_{n+1} - t_{n}, t_x^{n+1} | t_x^{n}, t) \rho_{n} (t_{n}, t_x^n | t_x^0, t) dt_n dt_x^n
\end{equation}
To simplify the integral equations, we use Bra-Ket notation and
thereafter, we derive the Laplace transform of the joint density by
\begin{equation}
  \tilde{\rho}_{n+1}(s , t_x^{n+1} | t_s^0, t)= \tilde{\rho}_{n} (s, t_x^n
  | t_x^0, t)| \tilde{\rho}(s , t_x^{n+1} | t_x^{n}, t) \rangle
\end{equation}
where $ \rho_{1}(t_{1} , t_s^{1} | t_x^0, t)= \rho(t_{1} , t_x^{1}
|t_x^0, t)$
% and
%
%\begin{equation}
%  \rho_{n+1}(t_{n+1} , t_x^{n+1} | t_x^0, t)= \int\int \rho( t_{n+1} - t_{n}, t%_x^{n+1} | t_x^{n}, t) \rho_{n} (t_{n}, t_x^n | t_x^0, t) dt_n dt_x^n
%\end{equation}
%
Next, we define the operator $\mathbf{P}_n(s|t)$, 
\begin{equation}
  \tilde{\rho}_{n}(s|t) =  \langle 1 \mid \mathbf{P}_n(s|t) \mid {\Pr}(t_s,t)  \rangle = \langle 1 \mid \mathbf{P}^n(s|t) \mid {\Pr}(t_x,t)  \rangle.
\label{ingenious2}\end{equation}
Now, by employing $\tilde{P}(1, s|t) = 1 / (\mu_1 s^2) [
\tilde{\rho}_{2}(s|t) - 2 \tilde{\rho}_{1}(s|t) +1]$ as in \cite{farkhooi_adaptation_2011}, we derive
\begin{equation}  
  \tilde{P}(n, s |t) = 1 / (\mu_1 s^2) [ \tilde{\rho}_{n+1}(s|t) - 2 \tilde{\rho}_{n}(s|t) + \tilde{\rho}_{n-1}(s|t)],
\label{count_n}\end{equation}
where $\tilde{P}(n, s | t)$ is the Laplace transform of the
probability density of observing $n$ events in a given time
window. Now we derive
\begin{equation} 
  \textstyle \tilde{\mathcal{A}}_{s|t} = \sum_k  \tilde{\rho}_{k}(s|t) =  \langle 1 \mid \mathbf{P}(s|t) / (\mathbf{I} -\mathbf{P}(s|t)) \mid {\Pr}(s|t)  \rangle
\label{auto}\end{equation}
where $\mathbf{I}$ is the identity operator. This equation represents
the Laplace transfom of the autocorrelation function.  Thus, the Fano
factor is $\tilde{J}_{s|t} = 1/ \mu_1 s^2 - (1 + 2 \tilde{
  \mathcal{A}}_{s|t})$ and the inverse Laplace transform is
\begin{equation}
  J_{T|t} =    1 + (2/T)  \int_0^T (T - u) \mathcal{A}(u|t) du - Tr_t ,
\label{Fano}\end{equation}
where $ \mathcal{A}(u|t) = \mathcal{L}^{-1}[\tilde{ \mathcal{A}}_{s|t}]$.

\subsection*{Computational model of insect olfactory coding}

Our model neuron is a general conductance-based integrate-and-fire
neuron with spike-frequency adaptation as it is proposed in Muller et
al.\cite{muller_spike-frequency_2007}. The model phenomenologically
captures a wide array of biophysical spike-frequency mechanisms such
as M-type current, afterhyperpolarization (AHP-current) and even slow
recovery from inactivation of the fast sodium
current\cite{muller_spike-frequency_2007}. The model neuron is also
known to perform high-pass filtering of the input frequencies
following the universal model of
adaptation\cite{benda_universal_2003}. Neuron parameters used follow
the Table.3 in Muller et al.\cite{muller_spike-frequency_2007} The
conductance model used for the static synapses between the neurons is
alpha-shaped with gamma distributed time constants from $\Gamma(2.5,
4)$ and $\Gamma(5,4)$ for excitatory and inhibitory synapses,
respectively.  All simulations were performed using the NEST
simulator\cite{gewaltig_nest_2007} version 2.0beta and the Pynest
interface.

The network connectivity is straight forward: each PN and LN receives
excitatory connections from 20\% randomly chosen
OSNs\cite{sachse_role_2002, chou_diversity_2010}.
Additionally, every PN receives input from 50\% randomly chosen
inhibitory LNs\cite{sachse_role_2002, chou_diversity_2010}.  In our
model the PNs do not excite one another and each PN output diverges to
50\% randomly chosen KCs\cite{szyszka_sparsening_2005,
  assisi_adaptive_2007,jortner_simple_2007}.

We tuned the simulated network by adjusting the synaptic weights to
achieve the same spontaneous firing rate as reported experimentally:
OSNs 15-25Hz\cite{nagel_biophysical_2011}, LNs
4-10Hz\cite{chou_diversity_2010}, PNs 3-10\cite{krofczik_rapid_2008}
and KCs ~ 0.3-1.0Hz\cite{ito_sparse_2008}.

\subsection*{Experimental methods}

Experiments were performed following the methods published in Szyszka
et al\cite{szyszka_sparsening_2005}. In summary, foraging honeybees
({\it Apis mellifera}) were caught at the entrance of the hive,
immobilized by chilling on ice, and fixed in a plexiglas chamber
before the head capsule was opened for dye injection. We retrogradely
stained clawed Kenyon cells (KC) of the median calyx, using the
calcium sensor FURA-2 dextran (Molecular Probes, Eugene, USA) with a
dye loaded glass electrode, which was pricked into KC axons projecting
to the ventral median part of the
$\alpha$-lobe\cite{szyszka_sparsening_2005}. After dye injection the
head capsule was closed, bees feed and kept in a dark humid chamber
for several hours.

The processing of imaging data was performed with custom written
routines in IDL (RSI, Boulder, CO, USA). In summary, changes in the
calcium concentration were measured as absolute changes of
fluorescence: a ratio was calculated from the light intensities
measured at 340nm and 380nm illumination and the background
fluorescence before odor onset was subtracted leading to $\Delta$F
with F = F340/F380.  Odor stimulation was preformed under a 20x
objective of the microscope, the naturally occurring plant odor
octanol (Sigma Aldrich, Germany), diluted 1:100 in paraffine oil
(FLUKA, Buchs, Switzerland), was delivered to both antennae of the bee
using a computer controlled, custom made olfactometer.  To this, odor
loaded air was injected into a permanent airstream resulting in a
further 1:10 dilution. Stimulus duration was 3 seconds if not
mentioned otherwise. The air was permanently exhausted.

For GABA blockage, a solution of 150 $\mu$l GABA receptor antagonist
dissolved in ringer for final concentration (10$^{-5}$M picrotoxin
(PTX, Sigma Aldrich, Germany) or 5x10$^{-4}$M CGP54626 (CGP, Tocris
Bioscience, USA)) was bath applied to the brain after pre-treatment
measurements. Measurements started 10 min after drug application.

\section*{Acknowledgments} We wish to thank Peter Latham and Yifat
Prut for helpful comments on this manuscript. Generous funding was
provided by the Bundesministerium f\"ur Bildung und Forschung (Grant
No.01GQ0941) to the Bernstein Focus Neuronal Basis of Learning (BFNL)
and by the Deutsche Forschungsgemeinschaft (DFG) to the Collaborative
Research Center for Theoretical Biology (SFB 618) and the DFG grant to
R.M. and A.F. (Me 365/31-1).

\bibliography{NNrefs}

%\newpage

%%%%%%%%%%%%%%%%%%%%%%%%%%%%%%%%%%%%%%%%%%%%%%%%%%%%%%%%%%
%%%%                FIGURES 
%%%%%%%%%%%%%%%%%%%%%%%%%%%%%%%%%%%%%%%%%%%%%%%%%%%%%%%%%%

%\section*{Figure Legends}

\end{document}